\documentstyle[epsfig,aps,floats,preprint,tighten]{revtex}

\def\beq{\begin{equation}}
\def\eeq{\end{equation}}
\def\qvec{{\bf q}}
\def\qhat{\hat{\bf q}}
\def\etal{{\it et al.}}
\def\epsbol{\mbox{\boldmath$\epsilon$}}
\def\sigbol{\mbox{\boldmath$\sigma$}}
\def\mpi{M_\pi}
\def\mn{m_{\scriptscriptstyle N}}
\def\N{{\scriptscriptstyle N}}
\def\A{{\scriptscriptstyle A}}

\begin{document}
\draft

\title{Spin polarisabilities of the nucleon at NLO in the chiral expansion}
\author{K. B. Vijaya Kumar\footnote{Permanent address: Department of Physics,
University of Mangalore, Mangalore 574 199, India\\ Electronic address:
kbvijayakumar@yahoo.com}, 
Judith A. McGovern\footnote{Electronic address: judith.mcgovern@man.ac.uk} 
and Michael C. Birse\footnote{Electronic address: mike.birse@man.ac.uk} }
\vskip 20pt
\address{Theoretical Physics Group, Department of Physics and Astronomy\\
University of Manchester, Manchester, M13 9PL, U.K.}
\nopagebreak
\maketitle
\begin{abstract}
We present a calculation of the fourth-order (NLO) contribution to 
spin-dependent Compton scattering in heavy-baryon chiral perturbation 
theory, and we give results for the four spin polarisabilities. No low-energy
constants, except for the anomalous magnetic moments of  the nucleon, enter
at this order. For forward scattering the fourth-order piece of the spin 
polarisability of the proton turns out to be almost twice the size of the
leading piece, with the opposite sign. This leads to the conclusion that no 
prediction can currently be made for this quantity.  For backward scattering
the fourth-order contribution is much smaller than the third-order  piece
which is dominated by the anomalous scattering, and so cannot explain the
discrepancy between the CPT result and the current best experimental
determination.
\end{abstract}


Compton scattering from the nucleon has recently been the subject of much
work, both experimental and theoretical. For the case of unpolarised protons
the experimental amplitude is well determined, and in good agreement with the
results of heavy-baryon chiral perturbation theory (HBCPT). However the
situation with regard to scattering  from polarised targets is less
satisfactory, not least because until very recently no direct measurements of
polarised Compton scattering had been attempted.

The usual notation for spin-dependent pieces of the scattering amplitude in
the Breit frame is, for incoming real photons of energy $\omega$ and
momentum $\qvec$ to outgoing real photons of the same energy energy and
momentum $\qvec'$,
\begin{eqnarray}
T&=&\epsilon'^\mu\Theta_{\mu\nu} \epsilon^\nu\nonumber\\
&=&i\sigbol\cdot(\epsbol'\times\epsbol)\,A_3(\omega,\theta)+
i\sigbol\cdot(\qhat'\times \qhat)\,\epsbol'\cdot\epsbol \,
             A_4(\omega,\theta)\nonumber\\
&&+\Bigl(i\sigbol\cdot(\epsbol'\times \qhat)\,\epsbol\cdot\qhat'-
i\sigbol\cdot(\epsbol\times \qhat')\,\epsbol'\cdot\qhat\Bigr)\,
             A_5(\omega,\theta)\nonumber\\
&&+\Bigl(i\sigbol\cdot(\epsbol'\times \qhat')\,\epsbol\cdot\qhat'-
i\sigbol\cdot(\epsbol\times \qhat)\,\epsbol'\cdot\qhat\Bigr)\,
             A_6(\omega,\theta)\nonumber\\
&&+i\sigbol\cdot(\qhat'\times \qhat)\,\epsbol'\cdot \qhat\,\epsbol\cdot\qhat'\,
             A_7(\omega,\theta)+\hbox{spin independent pieces}
\label{amp}
\end{eqnarray}
where hats indicate unit vectors.
By crossing symmetry the functions $A_i$ are odd in $\omega$.  The leading 
pieces in an expansion in powers of $\omega$ are given by low-energy
theorems\cite{LGG}, and the next terms contain the spin polarisabilities
$\gamma_i$:
\begin{eqnarray}
A_3(\omega,\theta)\!&=&\!
{e^2 \omega\over 2\mn^2}\Bigl(Q(Q+2\kappa)-(Q+\kappa)^2 \cos\theta\Bigr)
+4\pi\omega^3(\gamma_1+\gamma_5\cos\theta)\nonumber\\
&&-{e^2 Q(Q+2\kappa)\omega^3\over 8\mn^4}
+{\cal O}(\omega^5)\nonumber\\
A_4(\omega,\theta)\!&=&\! -{e^2\omega \over 2\mn^2 }(Q+\kappa)^2 
+4\pi\omega^3 \gamma_2 +{\cal O}(\omega^5)\nonumber\\
A_5(\omega,\theta)\!&=&\! {e^2 \omega\over 2\mn^2 }(Q+\kappa)^2 
+4\pi\omega^3\gamma_4 +{\cal O}(\omega^5)\nonumber\\
A_6(\omega,\theta)\!&=&\! -{e^2 \omega\over 2\mn^2 }Q(Q+\kappa)
+4\pi\omega^3\gamma_3 +{\cal O}(\omega^5)\nonumber\\
A_7(\omega,\theta)\!&=&\!{\cal O}(\omega^5)
\label{pol}
\end{eqnarray}
where the charge of nucleon is $Q=(1+\tau_3)/2$ and its anomalous magnetic 
moment is $\kappa=(\kappa_s+\kappa_v\tau_3)/2$.
Only four of the polarisabilities are independent since three are related by 
$\gamma_5+\gamma_2+2\gamma_4=0$.  The polarisabilities are also isospin 
dependent.

None of the polarisabilities has yet been measured directly.  The best 
estimates that exist at present are for forward scattering (where only $A_3$ 
contributes). The quantity  $4\pi f_2(0)$  is defined as ${\rm d}
A_3(\omega,0)/{\rm d}\omega$ at $\omega=0$, and according to the LET it depends
only on $\kappa^2$\cite{LGG}. The relevant polarisability  is
$\gamma_0=\gamma_1+\gamma_5$, which is related via a dispersion relation to
measurements at energies above the threshold for pion production, $\omega_0$:
\beq
\gamma_0={1\over 4\pi^2}
\int^\infty_{\omega_0}{\sigma_{-}(\omega)-\sigma_{+}(\omega)\over
\omega^3}d\omega, 
\eeq
where $\sigma_{\pm}$ are the parallel and antiparallel cross-sections for
photon absorption;  the related sum rule for the model-independent piece, due
to Gerasimov, Drell and Hearn\cite{GDH}, has the same form except that
$1/\omega$ replaces $1/\omega^3$.

Before direct data existed, the relevant cross-sections were estimated  from
multipole analyses of pion electroproduction  experiments \cite{karl,sand}. 
These showed significant discrepancies between the LET and the GDH sum rule
for the  difference of $f_2(0)$ for the proton and neutron, though the sum
was in good agreement.  Indeed even the sign of the difference was different. 
More recently, measurements have been made with MAMI at Mainz, for photon
energies between 200 and 800~MeV; the range will be extended downward to
140~MeV, and a future experiment at Bonn will extend it upwards to 3~GeV
\cite{thomas}.    The preliminary data from MAMI \cite{thomas} suggest a
continuing discrepancy between the LET and the sum rule for the proton,
though a smaller one than given by the multipole analysis. The most recent
analysis using electroproduction data, which pays particular attention to the
threshold region, also reduces the discrepancy somewhat \cite{krein1}.

The MAMI data does not currently go low enough in energy to give a reliable 
result for the spin polarisability, $\gamma_0$.  However electroproduction
data have also been  used to extract this quantity; Sandorfi \etal 
\cite{sand}  find  $\gamma_0^p=-1.3 \times 10^{-4}$~fm$^4$ and
$\gamma_0^n=-0.4 \times 10^{-4}$~fm$^4$, while the more recent analysis of
Drechsel \etal \cite{krein2} gives a rather smaller value of $\gamma_0^p=-0.6
\times 10^{-4}$~fm$^4$.  (We shall use units of $10^{-4}$~fm$^4$ for
polarisabilities from now on.)

 The spin polarisability has also been calculated in the framework of 
HBCPT: at lowest (third) order in the chiral expansion this gives
$\gamma_0=\alpha_{em}g_\A^2/(24 \pi^2 f_\pi^2 m_\pi^2)=4.51$ for both proton
and neutron, where the entire contribution comes from  $\pi N$ loops. The
effect of the $\Delta$ enters in counter-terms at fifth  order in standard
HBCPT, and has been estimated to be so large as to change the  sign
\cite{ber92}. The calculation has also been done in an extension of HBCPT
with an explicit  $\Delta$ by Hemmert \etal \cite{hemm}.  They find that the
principal effect is from the $\Delta$ pole, which contributes $-2.4$, with
the effect of $\pi\Delta$ loops being small, $-0.2$.  Clearly the next most
important contribution is likely to be the fourth-order $\pi N$ piece, and
this is the result which is presented  here.  The effects of the $\Delta$ at
NLO involve unknown parameters; one might hope that the loop pieces at least
will be small.

One other combination of the polarisabilities has been estimated from
low-energy data for Compton scattering from the proton by Tonnison
\etal\cite{ton}, namely that for backward scattering,
$\gamma_\pi=\gamma_1-\gamma_5$. This quantity is  dominated by the anomalous
$\pi^0$ exchange graph, which vanishes for forward scattering, but at third
and fourth order there are also pion loop contributions.  The experimental
value is $\gamma_\pi=-27.1$ with experimental  and theoretical errors of
about 10\% each. The HBCPT result of Hemmert \etal\  is $-36.7$, of which
$-43.5$ is the anomalous contribution, 4.6 is the $\pi N$  piece and  2.2
comes from including the $\Delta$ \cite{hemm}.

In the current paper we calculate the fourth-order contributions to all four 
polarisabilities.  Results for $\gamma_0$ have been previously presented by
Ji and collaborators \cite{osborne} and by ourselves \cite{kumar}.

\begin{figure}
  \begin{center} \mbox{\epsfig{file=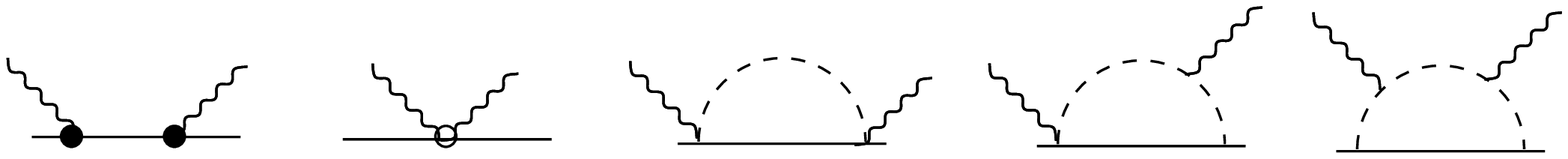,width=16truecm,angle=0}}
  \end{center}
{\bf Fig.~1:}
Diagrams which contribute to spin-dependent Compton scattering in the 
$\epsilon\cdot v=0$ gauge at LO. The solid dots are vertices
from ${\cal L}^{(2)}$ and the open circle is a vertex from ${\cal L}^{(3)}$.
\end{figure}

To calculate the spin-dependent scattering amplitude, we work in the gauge
$A_0=0$, or in the language of HBCPT, $v\cdot\epsilon=0$, where $v^\mu$ is
the unit vector which defines the nucleon rest frame.  Note that in this
gauge there is no lowest-order coupling of a photon to a nucleon; the 
coupling comes in only at second order. The Feynman vertex consists of two
pieces, one proportional to the charge current and one to the magnetic moment:
\beq {ie\over
2M}\Bigl(Q\epsilon\cdot(p_1+p_2)+2(Q+\kappa)[S\cdot\epsilon,S\cdot q]\Bigr).
\eeq This and all other vertices are taken from the review of Bernard  \etal
\cite{mei95}. 

At leading (third) order the fixed terms in the amplitudes $A_3$ to $A_6$
are reproduced, with $\kappa$ replaced by its bare value, by the combination
of the Born terms and the seagull diagram, which has a fixed coefficient in
the third-order Lagrangian \cite{ber92}. The loop diagrams of Fig.~1 have
contributions of order $\omega$ which  cancel and so do not affect the LET.
However they do give contributions to the polarisabilities.

At NLO, the diagrams which contribute are given in Fig.~2.  There can be no
seagulls at this order; since the NLO pieces of the $A_i(\omega)$ are of 
fourth chiral order and are odd in $\omega$, they will have expansions of
the form $e^2\omega\,(a\, m_\pi +b\,\omega^2/m_\pi+\ldots)$.  These non-analytic
powers of $m_\pi^2$ cannot be present in the basic couplings in the
Lagrangian, but can only be generated from loops.  It follows  that there
are no undetermined low-energy constants in the final amplitude.

The insertion on the nucleon propagator of Figs.~2a, b, h and i needs 
some explanation.  Denoting the external nucleon residual momentum by $p$,
the energy $v\cdot p$ starts at second chiral order with the mass shift and
kinetic energy. In contrast the space components of $p$ and all components
of the loop momentum $l$ and the photon  momentum are first order. The
propagator with an insertion  consists of both the second term in the
expansion of the lowest-order  propagator, $i/(v\cdot l+v\cdot p)$, in powers
of $v\cdot p/v\cdot l$, and also the insertions from ${\cal L}^{(2)}$. The
second-order mass shift and external kinetic energy cancel between the two.

\begin{figure}
  \begin{center} \mbox{\epsfig{file=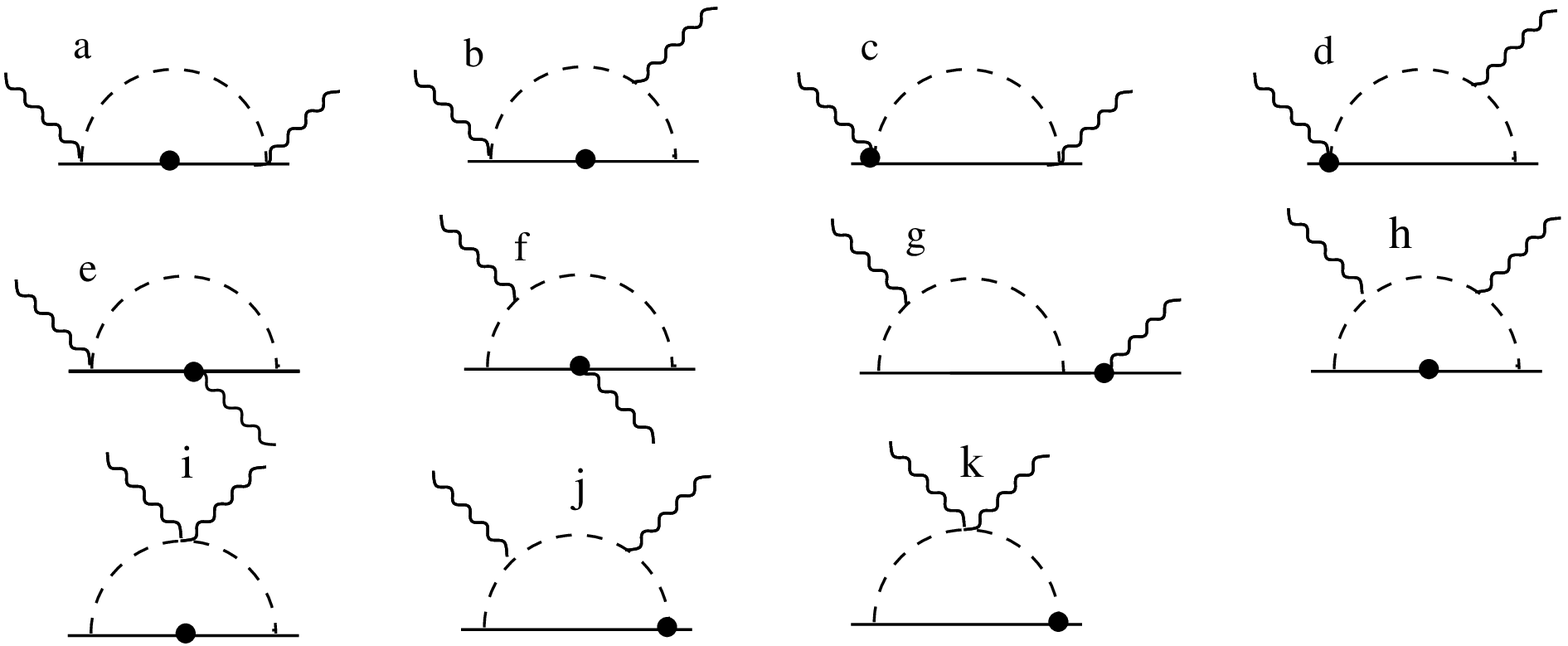,width=16truecm,angle=0}}
  \end{center}
{\bf Fig.~2:}
Diagrams which contribute to spin-dependent forward Compton
scattering in the $\epsilon\cdot v=0$ gauge at NLO. The solid dots 
are vertices from ${\cal L}^{(2)}$.
\end{figure}

To calculate the polarisabilities, it is sufficient to work in the Breit frame
in which the sum of the incoming and outgoing nucleon three-momenta is zero.
In the centre-of-mass frame, this is not the case, and at fourth chiral order
new terms appear in the amplitudes which were absent from the general form of
the  amplitude given by Hemmert \etal \cite{hemm}.  They can, however, easily 
be distinguished from the terms in which we are interested because they
appear to have the wrong crossing symmetry, contributing at order $\omega^2$
and $\omega^4$ in the expansions of the $A_i$.  These terms in fact arise from
boost corrections to the third-order contributions, the $\omega^2$ terms from
the LET pieces, and the $\omega^4$ terms from the third-order
polarisabilities.  There is also one other spin-dependent term---we need to
remember that a under a Lorentz boost,  Wigner rotations can generate
spin-dependent amplitudes from spin-independent  ones.  This effect is of
second order in the chiral expansion, in contrast to the other terms in a
boost  which are of first chiral order.  Since the Thomson term for
spin-independent Compton  scattering is of second order in HBCPT, through the
Wigner rotation it gives  rise to a fourth-order contribution to $A_4$ in any
but the Breit frame.

Returning to the Breit frame, only diagrams 2a-h contribute.  When the 
amplitudes are Taylor expanded, there are  contributions at order $\omega$ and
$\omega^3$.  The former do not violate the LETs, however; they simply provide
the pieces proportional to  $\delta \kappa_v= -g_\A^2 m_\pi M_\N/4 \pi
f_\pi^2$ necessary to satisfy the  LET's to fourth order. The contributions
of the various diagrams from Fig.~2 to the polarisabilities are given in
Table 1.  It can be seen that the requirement $\gamma_5+\gamma_2+2\gamma_4=0$
is satisfied, which provides a non-trivial check on the results.  

The loop contributions to the polarisabilities to NLO are then
\begin{eqnarray}
\gamma_1&=&{\alpha_{em}g_\A^2 \over 24\pi^2 f_\pi^2 m_\pi^2}\left[1 
-{\pi m_\pi \over 8 M_\N}\bigr(8+5\tau_3\bigr)\right]
\nonumber\\
\gamma_2&=&{\alpha_{em}g_\A^2 \over 48\pi^2 f_\pi^2 m_\pi^2}\left[1 
-{\pi m_\pi \over 4 M_\N}\bigr(8+\kappa_v+3(1+\kappa_s)\tau_3\bigr)\right]
\nonumber\\
\gamma_3&=&{\alpha_{em}g_\A^2 \over 96\pi^2 f_\pi^2 m_\pi^2}\left[1 
-{\pi m_\pi \over 4 M_\N}\bigr(6+\tau_3\bigr)\right]
\nonumber\\
\gamma_4&=&{\alpha_{em}g_\A^2 \over 96\pi^2 f_\pi^2 m_\pi^2}\left[-1 
+{\pi m_\pi \over 4 M_\N}\bigr(15+4\kappa_v+4(1+\kappa_s)\tau_3\bigr)\right]
\nonumber\\
\gamma_0&=&{\alpha_{em}g_\A^2 \over 24\pi^2 f_\pi^2 m_\pi^2}\left[1 
-{\pi m_\pi \over 8 M_\N}\bigr(15+3\kappa_v+(6+\kappa_s)\tau_3\bigr)\right]
\end{eqnarray} 
Although the subleading pieces have a factor of $m_\pi/M_\N$ compared with
the leading piece, the numerical coefficients are often large. The anomalous
magnetic moments are $\kappa_s=-0.12$ and   $\kappa_v=3.71$; with these
values the numerical results for the polarisabilities  to fourth order are
\begin{eqnarray}
\gamma_1&=&[-21.3]  +4.5 -(2.1 +1.3 \,\tau_3)\nonumber\\
\gamma_2&=&         2.3 - (3.1 + 0.7\,\tau_3)\nonumber\\
\gamma_3&=&[10.7]  +1.1 - (0.8 + 0.1\,\tau_3)\nonumber\\
\gamma_4&=&[-10.7]- 1.1  + (3.9 + 0.5\,\tau_3)\nonumber\\
\gamma_0&=&          4.5 -(6.9+1.5\,\tau_3)\nonumber\\
\gamma_\pi&=&[-42.7]+4.5 +(2.7 - 1.1\,\tau_3)
\end{eqnarray}
The term in square brackets, where it exists, is the third-order anomalous
contribution.  (The fourth-order anomalous contribution is proportional to the
difference in the incoming and outgoing photon energies, which vanishes in the
Breit frame.) 

The NLO contributions are disappointingly large, and call the
convergence of the expansion into question. While the 
fifth-order terms have also been estimated to be large\cite{ber92}, 
this is due to physics beyond pion-nucleon loops, namely the contribution
of the $\Delta$.  Our results show that even in the absence of the $\Delta$,
convergence of HBCPT for the polarisabilities has not yet been reached.
However the convergence is better
for $\gamma_\pi^p$, and so our results do not shed light on the 
large dicrepancy between the experimental and theoretical determinations.

\begin{figure}
\def\t3{$\tau_3$}
\begin{center}
\begin{tabular}{|c|c|c|c|c|c|} \hline
Diagram &$\gamma_1$ & $\gamma_5$ & $\gamma_2$  & $\gamma_4$ & $\gamma_3$
\\ \hline\hline
a  &--18  & --12 &   & & \\ \hline
b  &13  & 8 &   &4 & \\ \hline
c  &--12\t3&  &   & & \\ \hline
d  &4\t3  &  &   & & \\ \hline
e  &--3+3\t3  &  &   & & \\ \hline
f  &  &$-\mu_v+\mu_s\tau_3$  & $\mu_v-\mu_s\tau_3$  &  &$-(1-\tau_3)/2$ \\ 
\hline
g  &  &$-2(\mu_v+\mu_s\tau_3)$  & $-2(\mu_v+\mu_s\tau_3)$  
              &$2(\mu_v+\mu_s\tau_3)$ & $-1-\tau_3$\\ \hline
h  &  &  & --7  &$3/2$ &$-3/2$ \\ \hline
\end{tabular}
\end{center}                       
{\bf Table 1:}
Contributions to the polarisabilities from the diagrams of Fig.~2, in
units of $\alpha_{em}g_\A^2/(192\pi \mn\mpi f_\pi^2)$.  The polarisabilities 
are defined in Eq.~\ref{pol}.
\end{figure}

\medskip

After we completed this paper the J\"ulich group released their results  for
the fourth-order polarisabilities \cite{gellas}.  They choose not to include
the one-particle-reducible diagram of Fig.~2g in their definition of the 
polarisabilities. With the addition of that contribution, their results are in
agreement with ours.

\medskip

JMcG and MCB acknowledge the support of the UK EPSRC. VK held a Commonwealth
Fellowship while in Manchester.

\appendix
\section{Full amplitude}

The full amplitude in the Breit frame for diagrams 2a-2h are as follows.
The notation $t_i$ is used for the tensor structures which multiply the
amplitudes $A_i$.  
\begin{eqnarray}
T_a&=&{g^2 e^2\over 4\mn f_\pi^2}
\Bigl[\Bigl(m^2-\omega^2(1+\cos\theta)\Bigr)
{\partial J_0(\omega,m^2)\over\partial \omega}
-2\omega J_0(\omega,m^2)\Bigl]t_3 - (\omega\to-\omega)\nonumber\\
T_b&=&-{g^2 e^2\over 2\mn f_\pi^2}
\Bigl[{2m^2\over\omega}\Bigl(J_2'(\omega,m^2)-J_2'(0,m^2)\Bigr)t_3+
(1+\cos\theta){\partial J_2(\omega,m^2)\over\partial\omega}t_3
\nonumber\\
&&-\omega J_2'(\omega,m^2)t_5+
\omega \Bigl(t_5-2(1-\cos\theta)t_3\Bigr)\int_0^1\!dx J_2'(x \omega,m^2)\Bigr]
- (\omega\to-\omega)\nonumber\\
T_c&=&-{g^2 e^2\over 2\mn f_\pi^2}\tau_3\omega J_0(\omega,m^2)t_3
- (\omega\to-\omega)\nonumber\\
T_d&=&{g^2 e^2\over \mn f_\pi^2}\tau_3\omega \int_0^1\!dx J_2'(x\omega,m^2)t_3
- (\omega\to-\omega)\nonumber\\
T_e&=&{g^2 e^2\over 2\mn f_\pi^2}(1-\tau_3){1\over\omega}
\Big( J_2(\omega,m^2)-J_2(0,m^2)\Big)t_3
- (\omega\to-\omega)\nonumber\\
T_f&=&{g^2 e^2\over4 \mn f_\pi^2}\omega
\Bigl(2(\mu_v-\mu_s\tau_3)(t_3\cos\theta-t_4)+(1-\tau_3)t_6\Bigr)
\int_0^1\! dx (1-2x)J_2'(x\omega,m^2)
- (\omega\to-\omega)\nonumber\\
T_g&=&-{g^2 e^2\over 4\mn f_\pi^2}\omega
\Bigl(2(\mu_v+\mu_s\tau_3)(t_3\cos\theta+t_4-t_5)+(1+\tau_3)t_6\Bigr)
\int_0^1 \!dx J_2'(x\omega,m^2)
- (\omega\to-\omega)\nonumber\\
T_h&=&-{g^2 e^2\over \mn f_\pi^2}\omega^2\int_0^1 \!dy \int_0^{1-x}\!dx\Bigl[
\Big((7x-1)(t_6-t_5)+7(1-x-y)t_4\Bigr)
{\partial J_6''(\tilde\omega,m^2-xyt)\over\partial\tilde\omega}\nonumber\\
&&+\Bigl(2V(x,y,\theta)(xt_6-xt_5+(1-x-y)t_4)
\nonumber\\
&&\qquad\qquad\qquad\qquad-(1-x-y)(9xy-x-y)t_7\Big)
\omega^2{\partial J_2''(\tilde\omega,m^2-xyt)\over\partial\tilde\omega}
\nonumber\\
&&-xy(1-x-y)\omega^4V(x,y,\theta)t_7
{\partial J_0''(\tilde\omega,m^2-xyt)\over\partial\tilde\omega}\Big]
- (\omega\to-\omega)
\end{eqnarray}
where $\tilde\omega=(1-x-y)\omega$,
\beq
J_6(\omega,m^2)={1\over d+1}\Big((m^2-\omega^2)J_2(\omega,m^2)
-{\omega m^2\over d}\Delta_\pi\Bigr),
\eeq
$J_0(\omega,m^2)$, $J_2(\omega,m^2)$ and $\Delta_\pi$ have their usual
meanings, prime denotes differentiation with respect to $m^2$, and 
\beq
V(x,y,\theta)=(2xy-x-y+1)\cos\theta-x(1-x)-y(1-y).
\eeq


\begin{thebibliography}{99}

\bibitem{LGG} F. Low, Phys.\ Rev.\ {\bf 96} 1428 (1954); M. Gell-Mann and 
M. Goldberger, Phys.\ Rev.\ {\bf 96} 1433 (1954).

\bibitem{GDH} S. B. Gerasimov, Sov.\ J.\ Nucl.\ Phys.\ {\bf 2} 430 (1966);
S. D. Drell and A. C. Hearn, Phys.\ Rev.\ Lett.\ {\bf 16} 908 (1966).

\bibitem{karl} I. Karliner, Phys.\ Rev.\ {\bf D 7} 2717 (1973);

\bibitem{sand} A. M. Sandorfi, C. S. Whisnant and M. Khandaker, Phys.\ Rev.\
{\bf D 50}  R6681 (1994).

\bibitem{thomas} A. Thomas, Talk at PANIC 99, Uppsala, Sweden, June 1999.

\bibitem{krein1} D. Drechsel and G. Krein, Phys.\ Rev.\ {\bf D 58} 116009 
(1998).

\bibitem{krein2} D. Drechsel, G. Krein and O. Hanstein, Phys.\ Lett.\ 
{\bf B 420} 248 (1998).

\bibitem{ber92} V. Bernard, N. Kaiser, J. Kambor and U.-G. Mei\ss ner, Nucl.\
Phys.\  {\bf B 388} 315 (1992).

\bibitem{hemm} T. R. Hemmert, B. R. Holstein J. Kambor and G. Kn\"ochlein,
Phys.\ Rev.\ {\bf D 57} 5746 (1998).

\bibitem{ton} J. Tonnison, A. M. Sandorfi, S. Hoblit and A. M. Nathan,
Phys.\ Rev.\ Lett.\ 80 4382 (1998).

\bibitem{osborne} X. Ji, C-W.\ Kao and J. Osborne, Phys.\ Rev.\ {\bf D 61} 
074003 (2000).

\bibitem{kumar} K. B. V. Kumar, J. A. McGovern and M. C. Birse, 
{\tt hep-ph/9909442}.

\bibitem{mei95} V. Bernard, N. Kaiser and U.-G. Mei\ss ner, 
Int.\ J. Mod.\ Phys.\ E {\bf 4} 193 (1995). 

\bibitem{gellas} G. C. Gellas, T. R. Hemmert and U.-G. Mei\ss ner, 
{\tt nucl-th/0002027}.

\end{thebibliography}
\end{document}